# Clocking the anisotropic lattice dynamics of multi-walled carbon nanotubes by four-dimensional ultrafast transmission electron microscopy

Gaolong Cao[†], Shuaishuai Sun[†], Zhongwen Li, Huanfang Tian, HuaixinYang, and Jianqi Li[★]

Beijing National Laboratory for Condensed Matter Physics, Institute of Physics, Chinese Academy of Sciences, Beijing 100190, China,

Collaborative Innovation Center of Quantum Matter, Beijing 100084, China;

[†]These authors contributed equally to this work.  [★]e-mail: ljq@aphy.iphy.ac.cn

**Recent advances in the four-dimensional ultrafast transmission electron microscope (4D-UTEM) with combined spatial and temporal resolutions have made it possible to directly visualize structural dynamics of materials at the atomic level. Herein, we report on our development on a 4D-UTEM which can be operated properly on either the photo-emission or the thermionic mode. We demonstrate its ability to obtain sequences of snapshots with high spatial and temporal resolutions in the study of lattice dynamics of the multi-walled carbon nanotubes (MWCNTs). This investigation provides an atomic level description of remarkable anisotropic lattice dynamics at the picosecond timescales. Moreover, our UTEM measurements clearly reveal that distinguishable lattice relaxations appear in intra-tubular sheets on an ultrafast timescale of a few picoseconds and after then an evident lattice expansion along the radical direction. These anisotropic behaviors in the MWCNTs are considered arising from the variety of chemical bonding, i.e. the weak van der Waals bonding between the tubular planes and the strong covalent $sp^2$-hybridized bonds in the tubular sheets.**

Transmission electron microscopy (TEM) and electron diffraction, due to their fine spatial resolution, have played a very important role in the microstructural investigations of advanced materials. The experimental 2D and 3D micrographs in general could provide static structures with a remarkably high-spatial resolution. In particular, recent developments on the aberration-corrected TEM allow us to obtain images with a resolution on the sub-angstrom scale for crystalline samples[1, 2]. On the other hand, the frontier of TEM instrument developments in the past decade is increasingly focused on the in-situ and time-resolved experimentation as reviewed in previous literatures[3]. For instance, the developments of 4D-UTEM and relevant experimental techniques have been used broadly to study ultrafast structural dynamics in solids[4-6], and these investigations indeed have resulted in unprecedented insights into the lattice dynamics of solids undergoing structural evolution and phase transitions. In particular, recent studies on 4D-UTEM with single-electron pockets, space charge effects and their associated pulse broadening mechanisms have made remarkable progresses, reaching the femtosecond regime for imaging with electrons. Furthermore, the associated instrument developments are at present very rapid, frequently bringing new challenges and solutions to the new physical regimes in which these instruments operate[6, 7]. It is now possible to drive material processes with various combinations of heat, gaseous and focused lasers and to capture the evolution using either conventional video-rate acquisition or ultrafast pump-probe method[8, 9].

In this Article we report our recent works on developing a new 4D-UTEM and anisotropic lattice dynamics of MWCNTs investigated by this 4D-UTEM. Our study clearly reveals that the intra-sheet dynamic responses and inter-layer lattice relaxation of MWCNTs occur evidently in two different timescales. They are fundamentally in correlation with the specific tubular structure and chemical bonding behaviors, i.e. the strong covalent bond in tubular sheet and weak van der Waals inter-layer bonding along the radical direction. These results demonstrate that our UTEM can be used for the study of nanoscale energy transport, electron-phonon coupling and lattice relaxation in picosecond timescales.

## Results

**Set-up and characterization of UTEM.** In past three years, we have focused our attention on modifications of the conventional TEM gun for development of an UTEM at Institute of Physics, Chinese Academy of Sciences (IOP, CAS). Previously, it is noted that research groups in LLNS (Lawrence Livermore National Laboratory) [10] and Caltech (California Institute of Technology) [1, 11] have modified the TEM column structure by introducing an electron shift copper block between the anode and the condenser lens, moreover, a weak magnetic lens was also added in their UTEM to increase the number of electrons for time-resolved observations [10]. On the other hand, our project for developing UTEM started with modifications and improvements of an electron gun by re-designing the configuration, electronics and vacuum systems. It is expected that the modified gun can not only work well under photoelectron emission mode but also be operated as a conventional high-resolution TEM in a thermionic mode. Fig.1a shows a schematic representation of our conceptual design for the UTEM gun, in which we can introduce the ultrafast laser for driving photocathode from either laser-port 1 or laser-port 2. Actually the laser-port 1 is designed to introduce photo-emission laser in a field-emission gun as similarly discussed in ref.12. The laser-port 2 is designed for photo-emission associated with a thermionic gun. As a result of our three-year project and subsequent improvements, we have successfully developed a UTEM gun which works properly in a JEOL-2000EX for time-resolved imaging and conventional TEM observations as well.

Fig. 1b shows a photograph of the UTEM built at IOP, CAS. This microscope working with a modified electron gun can be operated for taking static and dynamic images as shown in the images Fig.1c and d, respectively. The relevant technical details on developments of this UTEM are illustrated partially in the supplementary materials [Fig. S1-S2] and will be reported elsewhere. It is known that spatial resolution in time-resolved observations depends evidently on numerous factors, such as the sample instability associated with optical excitation and space charge broadening. Although we can see lattice of the graphitized carbon under ideal conditions as shown in Fig.S3[4, 11], in general we can obtain time-resolved images with a spatial resolution of about 0.5nm as shown in Fig.1c, this fact suggests that ultrafast electron microscopy can be a key technique to characterize the microstructural phenomena of nanoscale dynamics, such as phase segregations and the anisotropic lattice relaxations.

In order to operate our UTEM with a high-temporal resolution, the concept of single-electron imaging and relevant experimental techniques are critically important to eliminate the coulomb repulsion between electrons within a femtosecond pocket[4]. In fact, our recent investigations demonstrated that single electron diffraction and imaging are practically possible in the UTEM mode. The inserted image of Fig.1b shows a single-electron diffraction (SED) pattern obtained from a MWCNT sample. This diffraction pattern is an accumulation of about $6\times10^7$ single-electron pulses (with the integration time about 10-mins), the presence of clearly diffraction spots and rings in this pattern demonstrates the ability of each individual electron to interfere with itself, as similarly discussed in ref.13. Moreover, it is also pointed out in previous literatures that as long as the number of electrons in each pulse is below the space-charge effect limit[4, 14], each pocket can actually have a few or tens of electrons, and the temporal resolution is still determined by the fs-optical pulse duration. Though precise measurement of the electron pulse durations under different operating conditions is a difficult work in the UTEM performance, we can efficiently estimate the duration of photo-electron pockets based on the UTEM parameters using the method as reported in ref. 14. e. g. when our UTEM is operated with the voltage of 160kV, the electron numbers per pocket should be less than 200 electrons to achieve an optimum temporal resolution of 1 ps or less, as clearly illustrated in supplementary materials (Fig. S2 and Table S1).

**Ultrafast lattice dynamics of MWCNTs.** We now turn to discuss the experimental results on the ultrafast lattice dynamics of MWCNTs. Because of the nearly one-dimensional electronic structure, electronic transport in metallic CNTs could occur ballistically over long lengths along the axial direction and phonons also propagate easily along the CNTs, the room temperature thermal conductivity for an individual CNT is reported to be greater than that of natural diamond and the basal plane of graphite [15, 16]. Importantly, our UTEM observations in MWCNTs directly reveal the presence of remarkable anisotropic dynamic properties arising from electron-phonon coupling, phonon relaxation and lattice strain. The MWCNT samples used in present study were arranged in well-aligned bunches or decussating patterns on the conventional TEM Cu-grids as shown in Fig. 2. As a result, this kind of MWCNT samples with a thickness of 20nm-30nm is suitable for the fs-laser excitation during UTEM observations. In order to clearly visualize the structural dynamics in MWCNTs, an ultrafast fs laser system with the duration of 300fs (wavelength of $\lambda$=520nm) was used to excite the charge-carriers in the nanotubes, then the excited electrons in the delocalized $\pi$ bond undergo a fast relaxation and subsequently result in visible changes in the lattice structure. These atomic-scale dynamics are carefully analyzed by UTEM observations at different time delays. Our experimental measurements clearly demonstrate the presence of visibly anisotropic dynamic nature along the CNT radical direction and within the tubular sheets, respectively.

One of the most notable structural features as revealed in UTEM observations is the appearance of visible lattice expansion along the tubular radical direction following the laser excitation. Fig. 3a shows two typical electron diffraction patterns obtained at the negative (-10ps) and positive (20ps) time delays for a fluence of 25mJ/cm$^2$. These diffraction patterns are taken from the textured MWCNTs with a decussating pattern (see Fig. 2), so it shows apparently the (002) diffraction spot (d=3.4Å) and (100) (d=2.45 Å) diffraction ring that can be used to study temporal evolutions associated with lattice relaxations. For facilitating the comparison, diffraction difference between these negative and positive patterns are also displayed in the right frame of Fig.3a, illustrating the visible shift for the (002) spots arising from thermal expansion along the radical direction. Careful analysis on the diffraction results obtained from a few well-characterized samples suggests that MWCNTs generally show visibly anisotropic dynamic responses in lattice spacing，i.e. the (002) and (004) peaks often shift their positions continuously with the increase of laser fluence, on the other hand, the (100) and (110) reflections show relatively small changes in their positions as also discussed in the following context. For the better view of these dynamic features, a few diffraction patterns were radical integrated to form a one-dimensional curve in which the position shifts for (002) spots can be clearly recognized. All data shown in Fig.3b are taken at the time delay of 20ps with laser fluences of F = 0, 25mJ/cm$^2$ and 50mJ/cm$^2$, respectively. Position changes for (002) diffractions are clearly indicated by a few arrows.

Fig. 3c shows radical expansion changes following with laser excitations at the positive time delay of 20ps. It is demonstrated that when the sample is excited under moderate fluences of below 120mW (~60mJ/cm$^2$), the interlayer expansion is proportional to the pump power and shows a linear increase in this regime. The maximum atomic motion, as initiated by ultrafast laser heating (~10$^{14}$ K /s$^{-1}$), is found to be around 0.10 angstroms along the radical direction. On the other hand, the radical expansion at higher values of fluence (e.g. larger than 150mW) shows up a notable nonlinear tendency towards saturation, this fact suggests that high laser excitation could induced defect structures and even damage the atomic structure of MWCNTs, as similarly discussed for graphite in the far-from-equilibrium regime[17] and also noted in our TEM examinations. In Fig.3c we also display the calculated lattice temperatures based on our experimental data and thermal expansion coefficient of $\alpha_{002}$ =2.6 × 10$^{-5}$ K$^{-1}$ [18], similar temperature rises can also be obtained from the pumping power and thermal parameters for MWCNTs.

Fig. 3d depicts the time dependence of the inter-planar space for three different laser powers corresponding to the fluence of 20mJ/cm$^2$, 30mJ/cm$^2$ and 40mJ/cm$^2$, respectively, where the heating laser beam diameter on the specimen is 60μm. It is clearly recognizable that the thermal expansion along the radical direction increases evidently in MWCNTs with the increase of pumping fluences. The characteristic time for lattice expansion from a monoexpential fit is 4±1ps, which is much faster than the reported data as measured using a laser duration of 16ps [19]. Careful examinations reveals that this radical expansion occurs at the time delays between 6 ps to 18ps, which could keep for hundreds of picoseconds depending on the thermal diffusion for the UTEM samples and then the recover to the pre-heating state. Moreover, one of the most critical issues concerned in present study is the correlation between this remarkable radical lattice expansion and the electron-phonon interaction in MWCNTs. We therefore have examined the diffraction intensity decays following with the laser excitation. It is commonly found that the diffraction intensity not only has a notable change accompanying with the inter-planar expansion but also shows up faster rate than the radical expansion as typically illustrated in Fig. 3e. Careful analysis suggests that this intensity decay results firstly from the electron–phonon coupling occurring chiefly within tubular sheets as discussed in flowing context.

**Anisotropic dynamic behaviors.** It is noted that in our study the (002) peak often shows notable diffusive feature due to lattice defects and complicated microstructure of MWCNTs. Therefore the lattice responses in the intra-tubular plane have also been carefully measured in our UTEM observations. One of the most remarkable features as revealed in the following text is the presence two notable dynamic processes following the laser excitation: the ultrafast dynamic response primarily occurs within the tubular sheets with the time constant of 2.5 ±1ps associated with a small lattice expansion along the tubular axial direction and another one occurs relatively slower yielding visible atomic motions along the radical direction.

In Fig. 4a, we plot the measured diffraction intensity for (100) ring as a function of the delay time, illustrating the structural change in the tubular sheets. The shown data were normalized to the average intensity obtained at negative times. After correcting for power losses, the pumping laser fluence is estimated to be around 30mJ/cm$^2$, it is found that the diffraction intensity is suppressed by 10% on the timescale of a few picoseconds as also demonstrate in another well-characterized sample. The atomic motion in correlation with the present phonon relaxation can be quantified by considering a time-dependent Debye-Waller factor. Now the vibration amplitude is time dependent and the diffraction intensity can be expressed as $I(t)/I_0 =$ $\exp(-s^2\delta u^2(t)/3)$, where I(t) is the intensity of a diffraction peak at a given time t after excitation, $I_0$ is the intensity before excitation, s is the scattering vector, and $\delta u^2(t)$ is the mean-square atomic displacement. Because (100) is proportional to the square of the atomic displacements, the intensity suppression of 10% corresponds roughly to 1.6% change in atomic displacements (0.03Å). Furthermore, it is known that the rate of diffraction change is essentially in correlation with the electron-phonon coupling in the tubular sheets. We therefore have analyzed our UTEM data in comparison with results of the charge relaxation as measured in ultrafast optical spectroscopy[20-22]. Our experimental data reveals a time constant of 2.5±1ps for phonon response, which is in good agreement with the charge relaxation time observed in ref. 20-22.

Another noteworthy feature of the data shown in Fig. 4, and also partially discussed in Fig. 3e, is the apparently different nature for ultrafast responses observed in the intra-tubular sheets and along the radical direction. Though the thermal-expansion coefficient in the intra-sheets for MWCNTs is very small: $\alpha_{100} = (0\pm0.1)\times10^{-6} K^{-1}$ at room temperature and $\alpha_{100} = 2.5\times10^{-6} K^{-1}$ at high temperatures[18], as a result, our UTEM measurements indeed can reveal this small intra-sheet expansion for laser fluence lager than 30mJ/cm$^2$. Fig. 4 shows temporal evolution of diffraction data obtained from a few well-characterized samples, it is recognizable that the first ultrafast response occurs between 0 and 6ps, as indicated by diffraction intensity decay (Fig.4a) and a small lattice expansion of about ~1$^0/_{00}$ along the tubular axis direction (Fig.4b), nevertheless the remarkable inter-planar expansion of about 1.5% as measured from the (002) spot begins at the time delay of about 6ps (Fig. 4c, solid vertical line), these facts directly demonstrate that the lattice dynamics contain two anisotropic transient processes. In MWCNTs, because photoexcited carriers are actually anisotropically and preferentially coupled to specific phonon modes, so the three-temperature picture and relevant theoretical model should be invoked to understand the nature of the laser induced heating of electrons and phonons[23, 24]. Importantly, according to the rate of diffraction changes and time scales for the specific electron-phonon coupling and anharmonic phonon-phonon interaction, our observations of ultrafast response in tubular sheets demonstrate the presence of a strong electron-phonon coupling and rapid lattice relaxations along the tubular-axial direction, which yields the fast decay of the diffraction intensity accompanying with phonon excitations. These remarkable structural features should be critical for understanding of specific electronic/thermal properties for this kind of one-dimensional structures.

On the other hand, the anisotropic dynamic feature are fundamentally in correlation with the variety of bonding types for MWCNTs. i.e. the weak van der Waals bonding between the annular layers and the strong covalent $sp^2$-hybridized bonds in each sheet. It is known that the thermal expansion in solid results essentially from the anharmonic oscillations of atoms. In present case, MWCNTs contain the weak van der Waals bond between tubular planes and strong covalent bonds in tubular sheets. It is known that the covalent bonds often yield a quick dynamic response and a relatively small lattice expansion. In general, the potential energy for carbon bonding can be written as $U=U(r)+\alpha\delta r^2+\beta\delta r^3+\ldots$, in which the anharmonic terms could be visibly larger for van de Waals bond than that for the covalent bond[25], Fig. 5 shows an schematic illustration on potential curves for both intra-tubular and inter-sheet interactions in the MWCNTs, in particular the asymmetric potential for the weak van der Waals bond is evidently exhibited. The average inter-sheet separation $r_c(T)$ could have a remarkable expansion following laser heating owing to the photo-excitation of anharmonic oscillation[25, 26]. Interestingly, it is noted that the large interlayer thermal-expansion as observed for MWCNTs were comparable to those of the graphite interlayer spacing similar with what observed in the x-rays diffraction[18], and this fact suggests that the MWCNTs are mostly adopted the scroll or mixed structures as illustrated in the inserted images of Fig.5 [18,19].

In conclusion, Our UTEM study clearly reveals a strong anisotropic nature of lattice dynamics for the MWCNTs, which are fundamentally in correlation with the specific tubular structure and chemical bonding behaviors, i.e. the strong covalent bond in tubular sheet and weak van der Waals inter-layer bonding along the radical direction. Importantly, our time-resolved structural analyses demonstrate that the intra-sheet dynamic responses and inter-layer lattice relaxation occur evidently in two different timescales. The intra-sheet dynamic feature is fundamentally in correlation with electron/phonon coupling and lattice relaxation along the tube-axial direction. These remarkable dynamic features should be critically important for understanding of the nanoscale energy conversion and phonon (electron) transport in this kind of one-dimensional tubular structures. With the development of UTEM, we can now study structural dynamics with high temporal and spatial resolution of advanced materials, we also plan to develop the aberration corrected UTEM equipped with STEM and EELS systems in next generation UTEMs, which can offer distinguished capability for probing the ultrafast structural dynamics in designed nanoscale materials and biological systems.

## Methods

The MWCNT samples used in present study were prepared using a method described by Huang et al.[27], then the MWCNT samples for UTEM observations were fabricated in bunches and decussating patterns on the TEM Cu-grids as typically shown in Fig.2a. Our high-resolution TEM investigations also demonstrated that the MWCNTs in each bunch are mostly well-aligned as clearly illustrating in Fig.2b and Fig.2c. These MWCNT films often have a thickness of 20nm to 30nm which is appropriate for the fs-laser excitation and UTEM studies. Moreover, the decussating pattern could yield clear diffraction spots with a fine signal-to-noise ratio which is essentially needed for analysis of ultrafast structural dynamics.

The UTEM built at our laboratory consists primarily of the femtosecond laser systems, UTEM gun, sample positioning chamber with the pumping laser-port, and CCD camera. The most significant new developments in this new machine are the modified UTEM gun which is capable for either time-resolved imaging or conventional high resolution TEM observations. There are two femtosecond laser systems with pulse durations of 100fs (80MHz) and 300fs (1-1MHz) are equipped with our UTEM for structural dynamic investigations. The time separation between laser pulses can be varied to allow complete heat dissipation in the examined specimen. We used the second femtosecond laser system for the ultrafast experiments of MWCNTs, which has relatively larger pulse energy and lower repetition rate. The output laser was split into two parts, one for pump laser after second harmonic generation (520 nm) and the other for probe laser after third harmonic generation (347nm). The pump laser was focused to 60μm by a convex lens with f=500mm and can obtain large pump fluence. The probe laser was focused to 100μm to meet the size of cathode.

Experimental measurements for MWCNTs are preformed based on the laser pump-electron probe method. Firstly, the MWCNT specimens were excited by the pump laser introduced into the TEM column from a fused silica laser window, which can generate a series of successive dynamic process for ultrafast observations. Another pulsed probe laser was introduced into TEM gun for producing electron pulses which were subsequently accelerated in the high voltage TEM gun (80kV-200kV) and then scattered by the excited MWCNT samples, as a results, the time-resolved electron diffraction pattern and transmission images can be directly obtained for addressing the structural dynamics in examined area. In our UTEM experiments, the selected area apertures with the diameter of 5μm-10μm were often used to get diffraction data from a homogeneous region in the MWCNTs.

## Acknowledgements

This work was supported by National Basic Research Program of China 973 Program (Grant Nos. 2011CBA00101, 2010CB923002, 2011CB921703, 2012CB821404), the Natural Science Foundation of China (Grant Nos. 11274368, 51272277, 11074292, 11004229, 11190022), and Chinese Academy of Sciences. Helpful discussions with Prof. Y. Zhu and Prof. H. Ding are gratefully acknowledged.

## Author contributions

J.L., X.Y., and F.T. conceived the project, designed and modified the UTEM electron gun. S.S, Z. L., and G.C. established the ultrafast laser system. G.C., F.T., and J.L. installed the UTEM electron gun and commissioned the UTEM system. Z. L., F.T., and X.Y. prepared the sample of MWCNTs. J.L., S.S and G.C. designed the ultrafast experiments and G.C., S.S, and Z. L. performed the experiments. S.S. and Z. L. analyzed the data. J.L. supervised the whole project. All authors contributed to the discussion of the results and to writing the manuscript.

## Competing financial interests

The authors declare no competing financial interests.

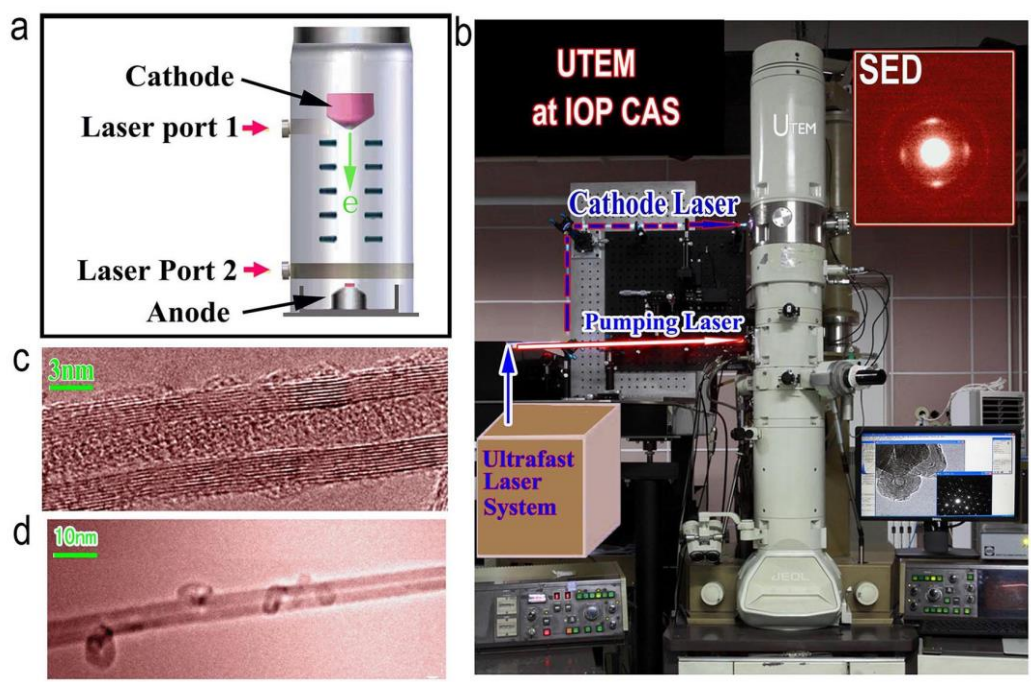

**Figure 1 Photograph of 4D UTEM at IOP，CAS, together with the relevant experimental data. a**, A conceptual design for an UTEM gun in which the ultrafast laser can be introduced for driving photocathode from either laser port 1 or laser port 2. **b**, A photograph of the UTEM at IOP, the inserted image is a single-electron diffraction (SED) pattern, demonstrating the ability of each individual electron to interfere with itself. **c** and **d,** Micrographs for a MWCNT taken with thermionic and UTEM mode, respectively.

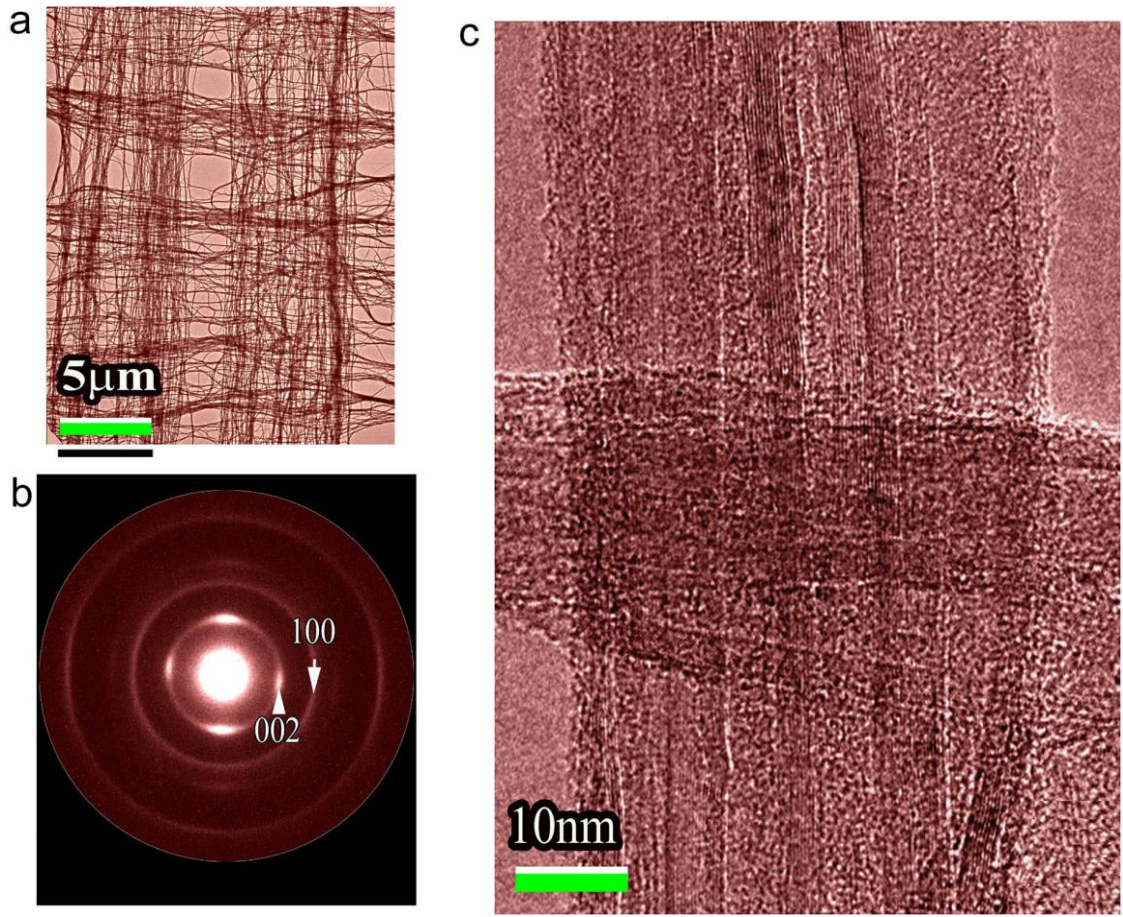

**Figure 2 Microstructural properties of MWCNT samples for UTEM observations. a,** The textured MWCNTs on a TEM Cu-grid. **b,** Electron diffraction pattern from the MWCNTs clearly show the (002), (004), (100) and (110) reflections. The decussating nantubes can yield a good signal-to-noise ratio in the ultrafast diffraction observations. **c**, High-resolution TEM image shows lattice structure of well-aligned nanotubes.

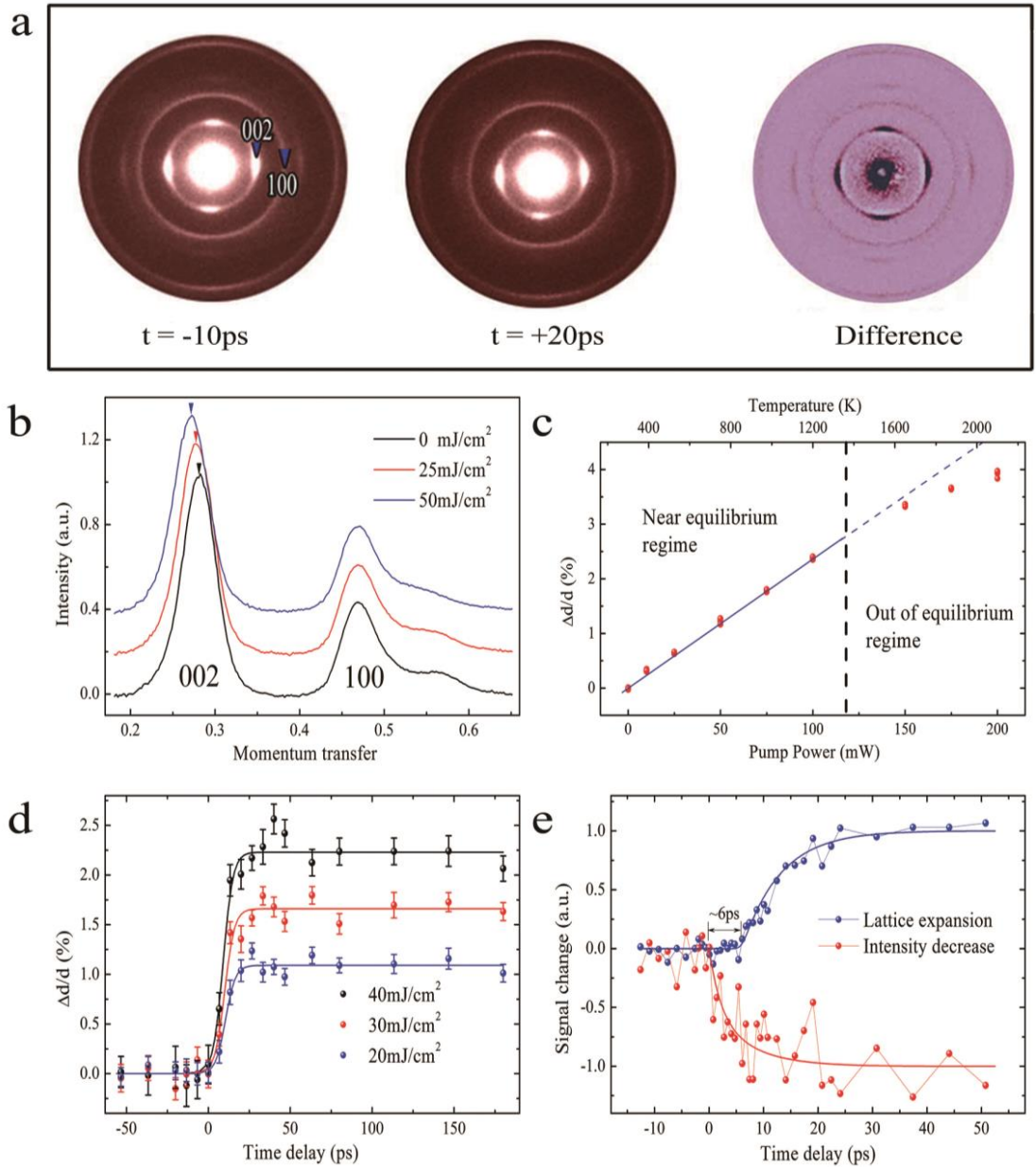

**Figure 3 Time-resolved experimental data for lattice dynamics along the radical direction of the MWCNTs. a**, Temporal frames of diffraction patterns for negative time delay (t = -10ps), positive time delay (t = 20ps), and diffraction difference clearly illustrating the shift of the (002) reflection. **b,** Radical integrated 1D diffraction curves for three different laser powers at the time delay of t = 20ps, showing the progressive move of (002) peak with the increase of pumping power. **c,** Fluence dependence of inter-planar expansion as measured at a time delay t = 20ps, dynamic anomalies occur for heating laser power larger than 120mW. **d,** Temporal evolution of lattice expansions along the radical direction with laser fluences of 20mJ/cm$^2$, 30 mJ/cm$^2$, 40mJ/cm$^2$, respectively. **e,** The rate of diffraction change revealing the (002) intensity decay and inter-planar expansion with time, the diffraction intensity has been normalized to the negative data. The notable intensity decay appears at about 6ps earlier than the notable lattice expansion.

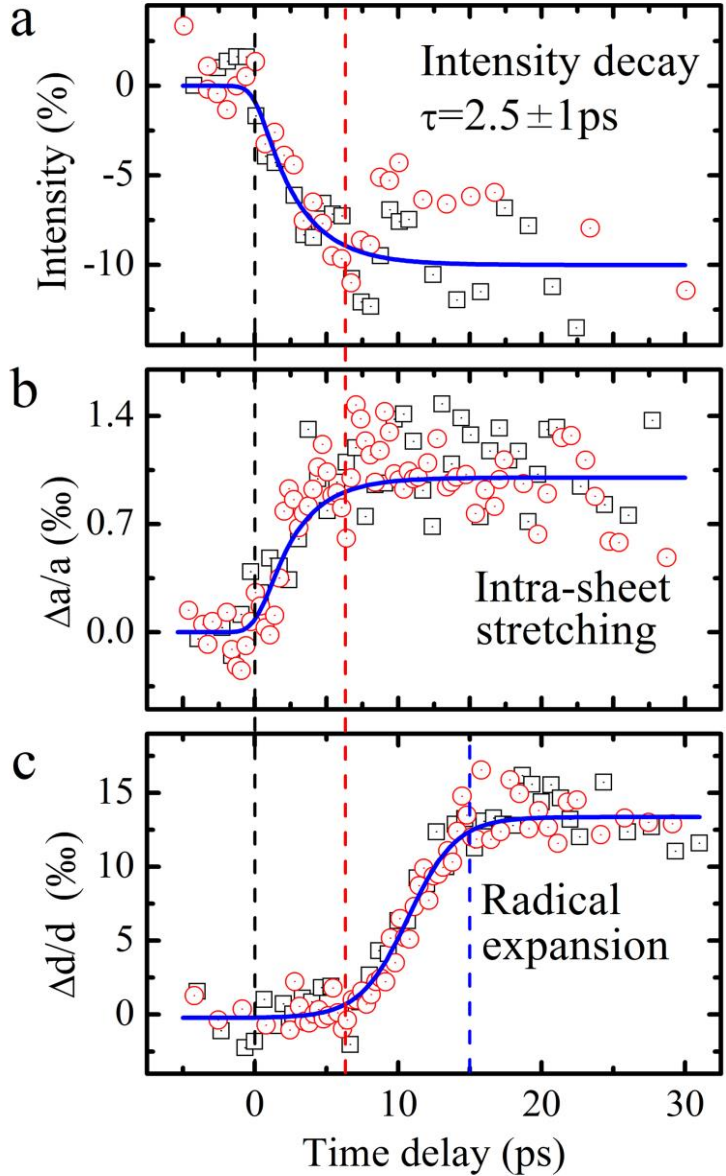

**Figure 4 Temporal evolution of the diffraction signals on MWCNTs taken within the intra-sheet，the typical data on inter-planar expansion along the radical direction is also shown for comparison. a,** Ultrafast decay of the (100) diffraction intensity with a time constant of 2.5±1ps, suggesting the presence of a strong electron-phonon coupling in the tubular structures. **b** and **c,** Comparison of the time dependence of lattice expansions as observed within the tubular sheets and along radical direction, demonstrating the anisotropic lattice relaxations in two different timescales.

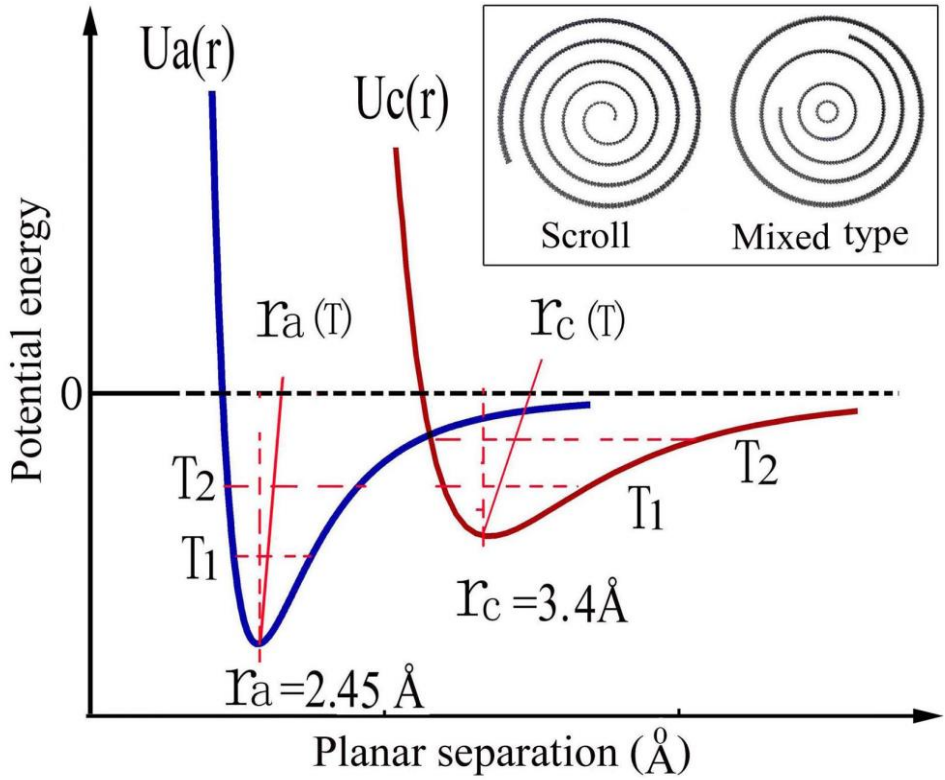

**Figure 5 Schematic patterns for potential functions of the strong covalent bond in tubular sheet ($U_a(r)$) and weak van der Waals inter-sheet bond ($U_c(r)$) along the radical direction.** At the room temperature we have $r_a$ = 2.45Å and $r_c$ = 3.4 Å, it is clearly illustrated that $r_c(T)$ could increase visibly resulting from the laser heating owing to the presence of apparent anharmonic feature in $U_c(r)$, This weak inter-layered interaction can be also written in the Lennard-Jones form, $U_c(r)=4\varepsilon[((\sigma/r)^{12}-(\sigma/r)^6]$ with $\varepsilon$ = 0.003eV and $\sigma$ = 0.34nm as used for graphite. Inserted images show two typical structural models for MWCNTs, so called the scroll type and mixed MWCNTs, they both could yield remarkable inter-planar expansion as observed in experiments.

Supplementary information

Our development project for an ultrafast transmission electron microscope (UTEM) started with modifications of the configuration, electronics and vacuum systems based on a 200kV conventional TEM electron gun. It is expected that the modified gun can not only well work under photoelectronic emission mode but also be operated as a conventional high-resolution TEM in a thermionic (or a field emission) mode. In Fig.1, we show a photograph of the UTEM at IOP. Presently, it works properly for time-resolved imaging and conventional TEM observations as well.

**1. Cathodes for the UTEM gun**

During the developments of UTEM, it is commonly noted that the fundamental properties of the photo-cathodes play critical role for obtaining high-quality time-resolved images, so primarily we need good cathode sources for the UTEM gun assembly. Though we have measured and analyzed the experimental data obtained from a few good candidates in past years (e.g. $LaB_6$, Ta, $CeB_6$, $SrB_6$), $LaB_6$ is the only thermionic source commonly used in conventional high-resolution TEM observations, therefore, in this paper we will main report our experimental data on the implementation of photoelectric emission TEM gun on a JEOL-2000EX microscope with the $LaB_6$ cathode plates.

**2. Time-zero determination**

It is known that there are a few fundamental issues have to be addressed for time-resolved UTEM observations, such as, the probe-pump synchronization, the space charge broadening, the source brightness and the electron coherence.

In order to perform pump-probe measurements on the UTEM mode, we firstly have adjusted our laser/TEM system and calibrate the synchronization between pumping fs laser and electron-pulse probe within subpicosecond precision. Actually, there are a few efficient methods that could well synchronize fs laser and electron pockets as used for UED and UEM systems. In our experiments, we found that it is convenient to use the untrashort electron pulses to image electron cloud of the surface plasma emitted from a metal surface following ultrafast laser illumination. Similar method is extensively discussed for imaging the electron shadow on a Cu TEM grid following fs-laser excitation [1]. Moreover, this kind of measurements is also appreciated for the discussion of space charge effects and time resolution under specific conditions, Fig.S1 shows the time resolved experimental data to determine the position of t = 0.

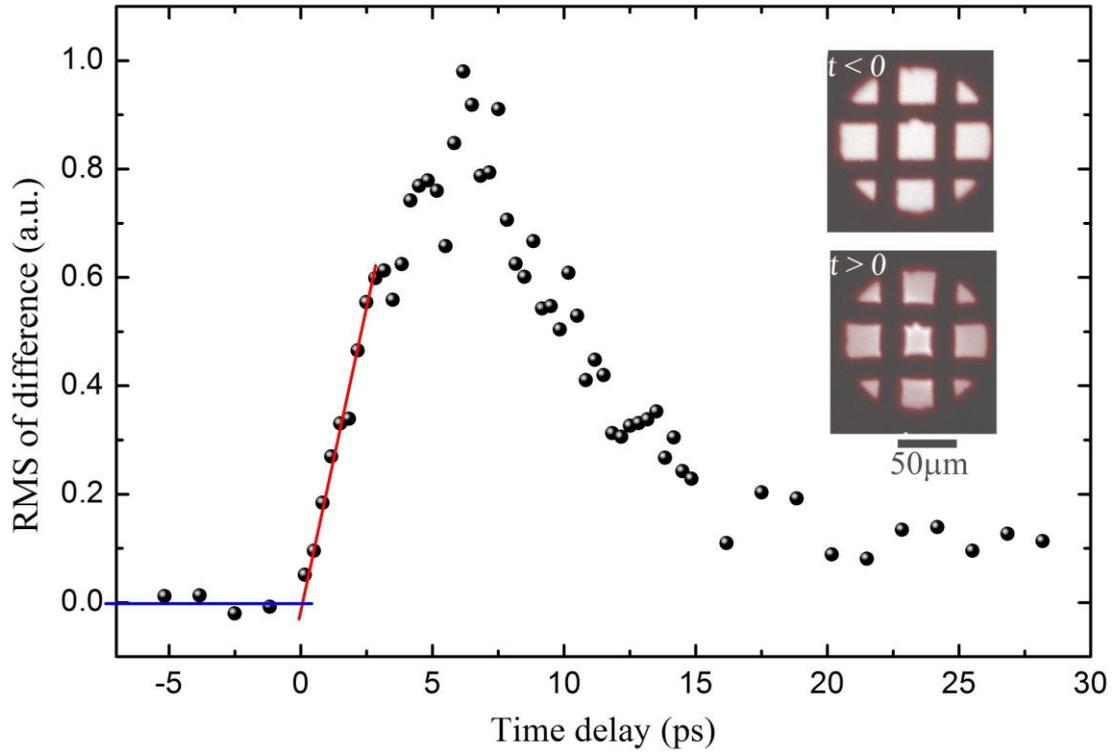

**Figure S1 Temporal evolution of electron image for a Cu-grid excited by the fs laser pulses ($\lambda$ = 520nm, F = 100mJ/cm$^2$).** Normalized changes as measured from the root mean square (RMS) for signal difference as a function of the time delay. The copper TEM grid is excited by the pump laser at the center area to generate plasma cloud which pushes the probe electron pulses away from the Cu grid following photo-excitation. Time zero can be determined at the onset of plasma cloud formation. Inserted images shows typically the time resolved Cu grid images (400 mesh) before and after the arrival of pumping laser, in which the evident effects of excited surface plasma on the probe electron pulses can clearly recognized at the center region of the image.

### 3. Space charge broadening effects

It is known that temporal resolution of UTEM images are essentially limited by space charge effects, and certain ultrashort processes cannot be observed. In order to study the space charge broadening effects and duration of the photo electron pulse resulting from the alteration of cathode laser, we have performed our measurements for the probe laser between 10μJ/cm$^2$ and 10mJ/cm$^2$. Indeed we can see visible changes of the plasma cloud images, careful analysis suggests that the probe electron pockets are strongly broaden by space charge effects for the laser fluences larger than F=1mJ/cm$^2$. In this project, we have used two femtosecond laser systems with the pulse durations of 100fs (λ=355nm, repetition rate 80MHz) and 300fs (λ=347nm, repetition rate 1MHz), respectively, they both can work properly for producing probe electron pockets.

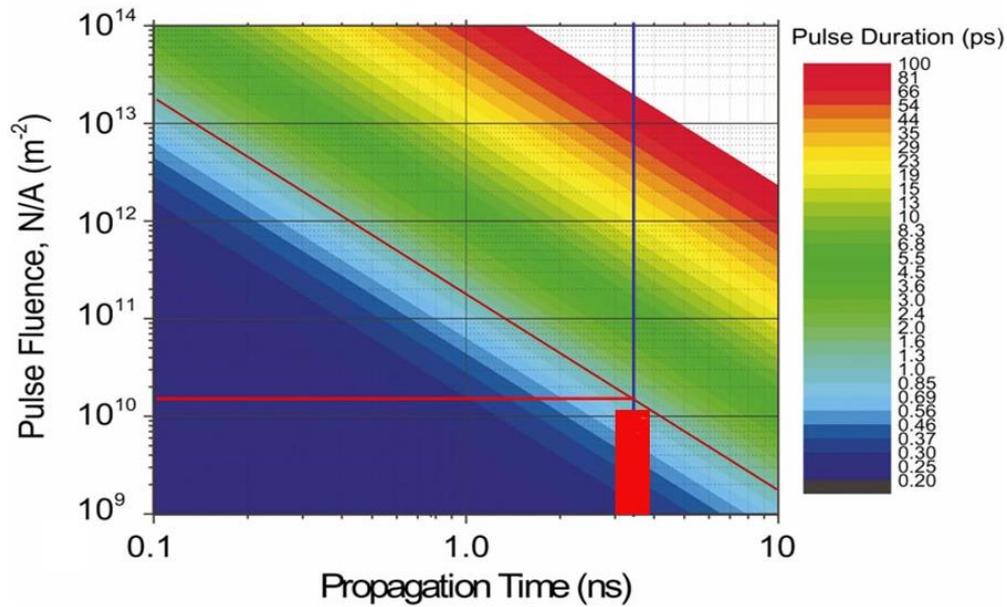

**Figure S2 Temporal broadening arising from space charge effects for femtosecond electron pulses**. This map illustrates clearly about pulse broadening effects for a given electron number density[2]. Our UTEM time-resolved experiments are mainly performed in the red boxed area with the sub-picosecond pulse durations.

**Table S1** Theoretical estimation for electron numbers in the each electron pocket for getting the subpicosecond time-resolution in UTEM at IOP, these calculations are based on the method reported ref.2 and the relevant UTEM parameters used in our experiments.

| Cathode Diameter | Operation voltage/Electron Number per Pulse | | |
|:---:|:---:|:---:|:---:|
| | 120kV | 160kV | 200kV |
| 16μm | 2.56 | 3.84 | 5.12 |
| 50μm | 25 | 37.5 | 50 |
| 100μm | 100 | **150** | 200 |

## 4. Micrograph for carbon nano-particles obtained in the UTEM mode

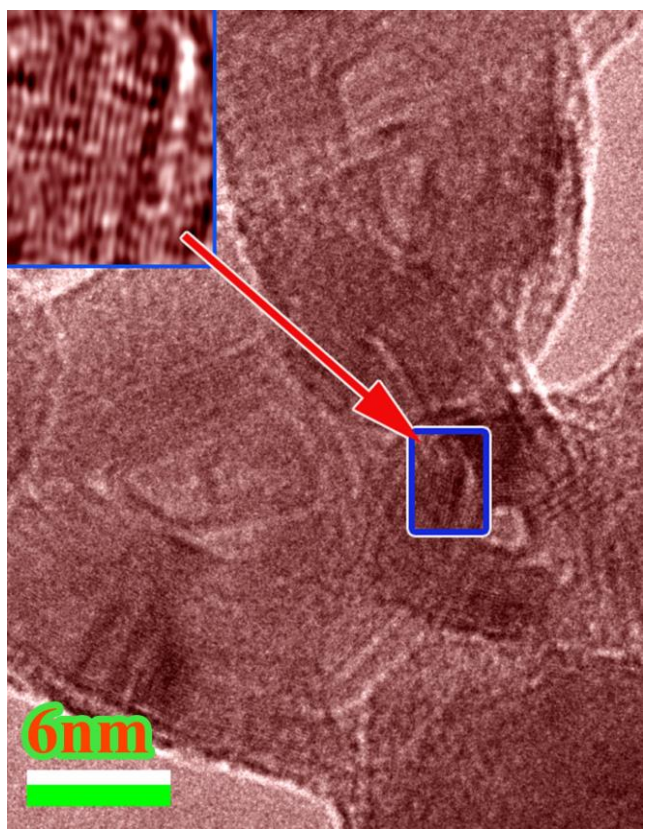

**Figure S3 Micrograph for carbon nano-particles obtained in the UTEM mode**. The lattice fringe d = 0.34nm can be observed at the center area (Cathode laser: Wave length λ=355nm, repetition rate: 80MHz).